\documentclass[printer]{aa}  
\usepackage{graphicx}
\usepackage{txfonts}
\usepackage{threeparttable}
\usepackage{graphicx}
\usepackage{amssymb}
\usepackage{amsmath}
\usepackage{algorithm} 
\usepackage{algpseudocode}
\usepackage[normalem]{ulem}

\newcommand{\RM} [1]{\mathrm{#1}}

\newcommand{\EQ}[1]{equation~(\ref{eq:#1})}
\newcommand{\FIG}[1]{Fig.~\ref{fig:#1}}

\newcommand{\SEC}[1]{Section~\ref{sec:#1}}
\newcommand{\APP}[1]{APPENDIX~\ref{sec:#1}}
\newcommand{\AG}[1]{Algorithm~\ref{ag:#1}}

\newcommand{\cmpc}{$\RM{cm}^{-3}\ \RM{pc}$}

\begin{document} 

   \title{PulsarX: a new pulsar searching package -I. A high performance folding program for pulsar surveys}


   \author{Yunpeng Men\inst{1}
          \and
          Ewan Barr\inst{1}
          \and
          C. J. Clark\inst{2,3}
          \and
          Emma Carli\inst{4}
          \and
          Gregory Desvignes\inst{1}
          }

   \institute{Max-Planck-Institut f\"{u}r Radioastronomie, Auf dem H\"{u}gel 69, D-53121 Bonn, Germany\\
              \email{ypmen@mpifr-bonn.mpg.de}
         \and
             Max Planck Institute for Gravitational Physics (Albert Einstein Institute), D-30167 Hannover, Germany
        \and
             Leibniz Universit\"{a}t Hannover, D-30167 Hannover, Germany
        \and
             Jodrell Bank Centre for Astrophysics, Department of Physics and Astronomy, The University of Manchester, Manchester M13 9PL, UK 
             }

   \date{Received XX XX, XXXX; accepted XX XX, XXXX}

 
  \abstract
   {Pulsar surveys with modern radio telescopes are becoming increasingly computationally demanding. This is particularly true for wide field-of-view pulsar surveys with radio interferometers, and those conducted in real or quasi-real time. These demands result in data analysis bottlenecks that can limit the parameter space covered by the surveys and diminish their scientific return.}
   {In this paper, we address the computational challenge of `candidate folding' in pulsar searching, presenting a novel, efficient approach designed to optimise the simultaneous folding of large numbers of pulsar candidates. We provide a complete folding pipeline appropriate for large-scale pulsar surveys including radio frequency interference (RFI) mitigation, dedispersion, folding and parameter optimization.}
   {By leveraging the Fast Discrete Dispersion Measure Transform (FDMT) algorithm proposed by Zackay et al. (2017), we have developed an optimized, and cache-friendly implementation that we term the pruned FDMT (pFDMT). This implementation is specifically designed for candidate folding scenarios where the candidates are broadly distributed in dispersion measure (DM) space. The pFDMT approach efficiently reuses intermediate processing results and prunes the unused computation paths, resulting in a significant reduction in arithmetic operations. In addition, we propose a novel folding algorithm based on the Tikhonov-regularised least squares method (TLSM) that can improve the time resolution of the pulsar profile.}
   {We present the performance of its real-world application as an integral part of two major pulsar search projects conducted with the MeerKAT telescope: the MPIfR-MeerKAT Galactic Plane Survey (MMGPS) and the Transients and Pulsars with MeerKAT (TRAPUM) project. In our processing, for approximately 500 candidates, the theoretical number of dedispersion operations can be reduced by a factor of around 50 when compared to brute-force dedispersion, which scales with the number of candidates.}
   {}


   \maketitle
%

\section{Introduction}

Pulsars are compact stars which emit pulsed radiation at their rotational period  \citep[[e.g.][]{Hewish1968Nat}. The remarkable rotational stability of pulsars allows them to be treated as precise clocks, providing a natural tool for the measurement of astrophysical phenomena \citep[e.g.][]{Verbiest2008ApJ}. Pulsars in binary systems may be used to test the limits of general relativity \citep[e.g.][]{Kramer2021PhR} or the equation of state of supra-nuclear matter \citep[e.g.][]{Demorest2010Nat, Antoniadis2013Sci}, while an array of pulsars distributed across the sky offers a natural detector for low-frequency gravitational waves \citep[e.g.][]{Antoniadis2023arXiv, Agazie2023ApJ, Reardon2023ApJ, Xu2023RAA}. Furthermore, in the case of radio pulsars, the broadband radio emission is altered by its propagation through the interstellar medium (ISM), allowing inference of properties such as the Galactic free electron density \citep[e.g.][]{Cordes2002astroph, Yao2017ApJ} and magnetic field \citep[e.g.][]{Han2018ApJS}.

Since the discovery of the first pulsar \citep[e.g.][]{Hewish1968Nat}, more than 3000 sources have been discovered through pulsar searches, including Galactic plane and all-sky surveys \citep[e.g.][]{Large1968Nat, Manchester1996MNRAS, Cordes2006ApJ, Keith2010MNRAS, Barr2013MNRAS, Stovall2014ApJ, Sanidas2019AA, Han2021RAA, Padmanabh2023arXiv}, globular clusters searches \citep[e.g.][]{Ransom2005ASPC, Hessels2007ApJ, Possenti2001astroph, Pan2021ApJ, Ridolfi2021MNRAS, Ridolfi2022A&A} and targeted observations of high-energy sources \citep[e.g.][]{Ransom2011ApJ, Cognard2011ApJ, Keith2011MNRAS, Kerr2012ApJ, Camilo2015ApJ, Barr2013bMNRAS, Cromartie2016ApJ, Wang2021SCPMA, Bhattacharyya2021ApJ, Clark2023MNRAS}. 

In these surveys, {\sc{sigproc}} \citep{Lorimer2011ascl} and {\sc{presto}} \citep{Ransom2002AJ} have been two of the most popular software packages used for pulsar searching, both of which are CPU-based. The pulsar searching pipeline usually consists of several key stages: (1) Radio frequency interference (RFI) mitigation is performed to remove non-astrophysical signals from the raw data; (2) Dedispersion is performed to correct for the frequency dependent delay caused by the propagation of the radio signal through the ISM; (3) Acceleration searching is performed to detect signals with periods that are changing as a function of time due to binary motion; (4) Candidate folding is performed to extract the time- and frequency-resolved pulse profile of each detected signal. To improve performance, GPU-based software has also been developed to speed up the acceleration search \citep[e.g.][]{Allen2013ApJ, Dimoudi2018ApJS, Barr2020ascl}, e.g. {\sc{peasoup}\footnote{https://github.com/ewanbarr/peasoup.git}}. However, as presented in \cite{Lyon2016MNRAS}, the number of candidates produced by contemporary pulsar surveys is increasing, which is the result of improving survey technical specifications, such as time resolution and bandwidth. Furthermore, applying a signal-to-noise ratio (S/N) filter to limit the number of candidates is ineffective, and advanced candidate selection mechanisms are required, such as pulsar candidate classifiers \citep[e.g.][]{Lee2013MNRAS, Zhu2014ApJ, Balakrishnan2021MNRAS}. These classifiers work on post-folding data, which requires a larger number of candidates for folding initially. Therefore, a more efficient candidate folding process is helpful to deal with the increasing number of candidates. In this work, we have developed a new high performance folding program to address this performance issue, which is part of a new developing pulsar searching package, {\sc PulsarX}\footnote{https://github.com/ypmen/PulsarX}.

The algorithms used in our new folding program are presented in \SEC{Algorithm}. We show the benchmarks in \SEC{Benchmark}. The application of {\sc PulsarX} in the MPIfR-MeerKAT Galactic Plane Survey of L-band \cite[MMGPS-L;][]{Padmanabh2023arXiv} is discussed in \SEC{Discussions} and the conclusions are summarized in \SEC{Conclusions}.

\section{Algorithm}
\label{sec:Algorithm}
The purpose of the folding pipeline is to generate data cubes that can be used for visualisation or for classification by machine learning classifiers. Candidates from the acceleration search are parameterized by a spin frequency $\nu$, spin frequency derivative ${\dot{\nu}}$ and dispersion measure (DM), as defined in \EQ{dispersion_delay}. In the folding process, the time stamp $t$ of each sample in the input time series is transformed to spin phase $\phi$ which is predicted from $\nu$ and ${\dot{\nu}}$, i.e.
\begin{equation}
     \phi = \phi_0 + \nu t + \frac{1}{2} \dot{\nu} t^2\,,
     \label{eq:phase}
\end{equation}
where $\phi_0$ is the reference phase at $t=0$. By integrating data samples over a discrete range of phase values we obtain the signal power as a function of spin phase otherwise known as the pulsar's `profile'. To classify a candidate, it is useful to view the profiles in different frequency sub-bands and time sub-integrations. Therefore, we perform the folding process over discrete, contiguous ranges of time and frequency. The resulting collection of folded pulse profiles is referred to as an `archive'. The reference phase of all profiles in the archive can be adjusted to account for the difference between true and used values of DM, $\nu$ and ${\dot{\nu}}$ of the candidate. The S/N of the candidate, summed over all profiles, will be maximised as these values approach their true values.

\begin{figure}
    \centering
    \includegraphics[width=\columnwidth]{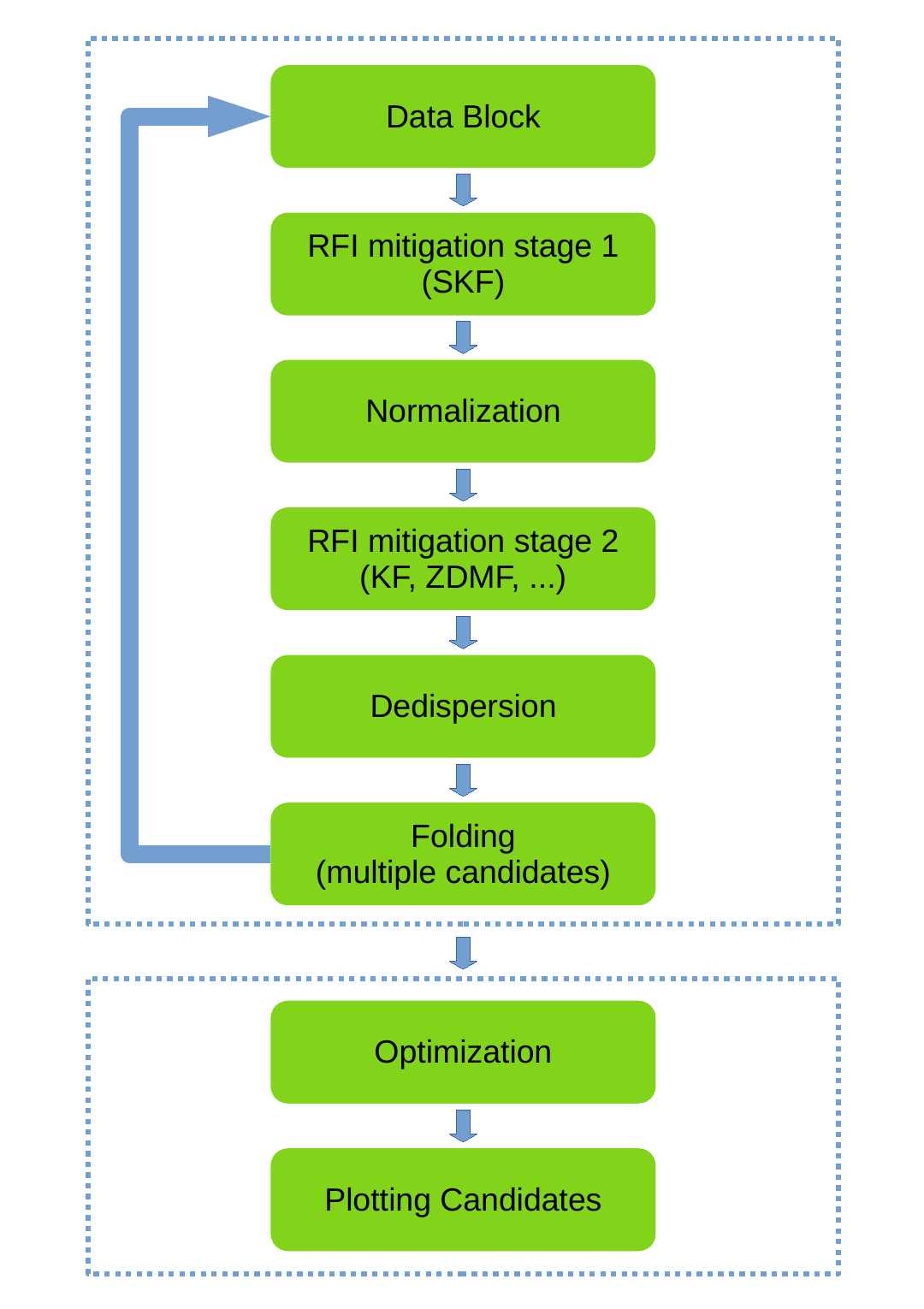}
    \caption{Data flow diagram of the candidate folding pipeline. The raw data is divided into blocks, and for each block, the following steps are performed: RFI mitigation, normalization, dedispersion and folding. Once all the data blocks have been folded, parameter optimization and candidate plotting are carried out for each candidate.}
    \label{fig:data_flow_diagram}
\end{figure}

Because the raw data is often contaminated by RFI signals, it is necessary to perform RFI mitigation before the folding process. The data is then dedispersed (see Section \ref{sec:dedispersion}) into frequency sub-bands after RFI mitigation. These processes will be discussed in the following sections, and the data flow diagram is shown in \FIG{data_flow_diagram}.

\subsection{RFI mitigation}
\begin{figure*}
    \centering
    \includegraphics[width=0.8\columnwidth]{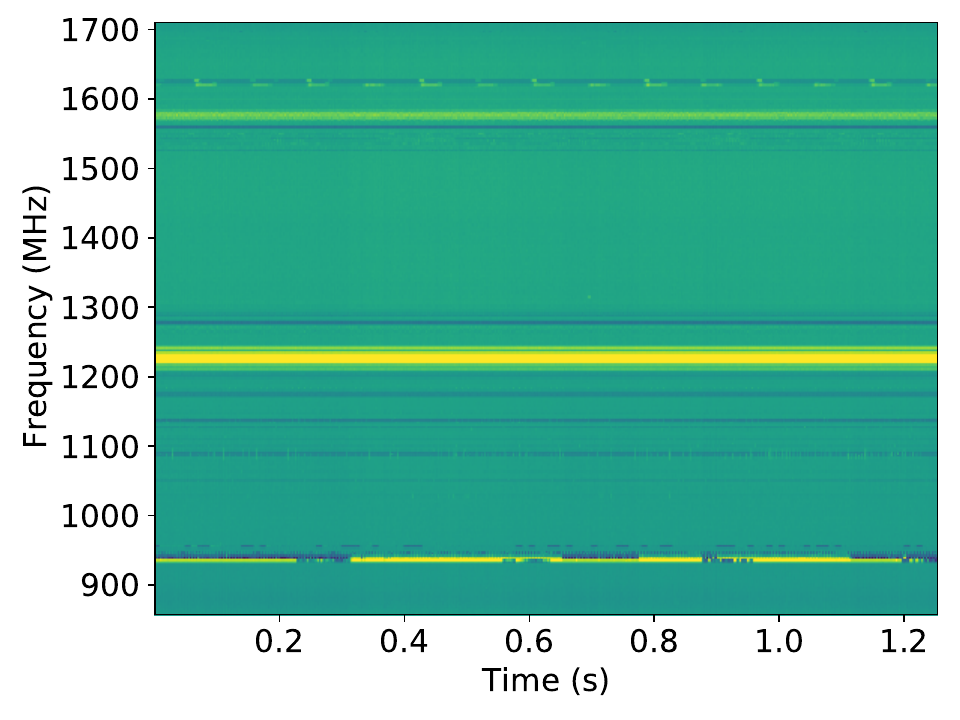}
    \includegraphics[width=0.8\columnwidth]{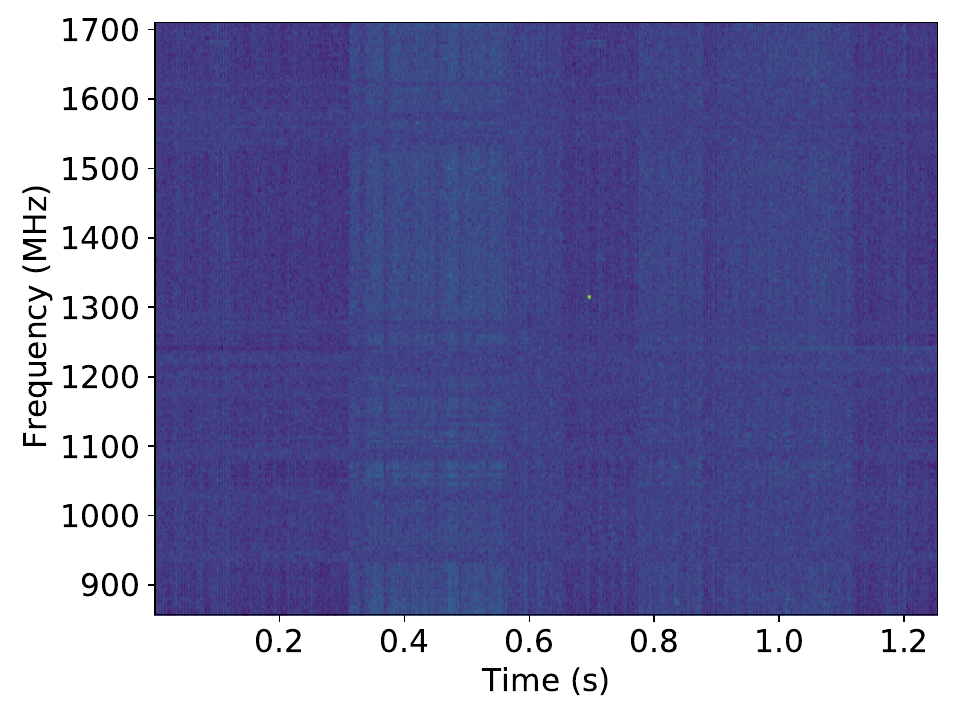}
    \includegraphics[width=0.8\columnwidth]{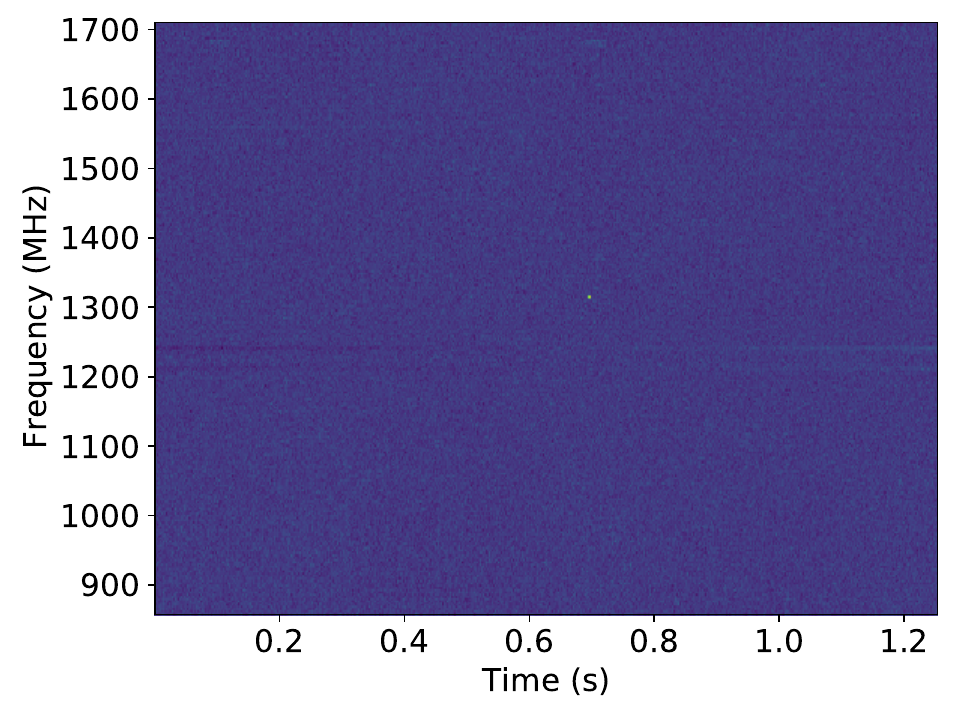}
    \includegraphics[width=0.8\columnwidth]{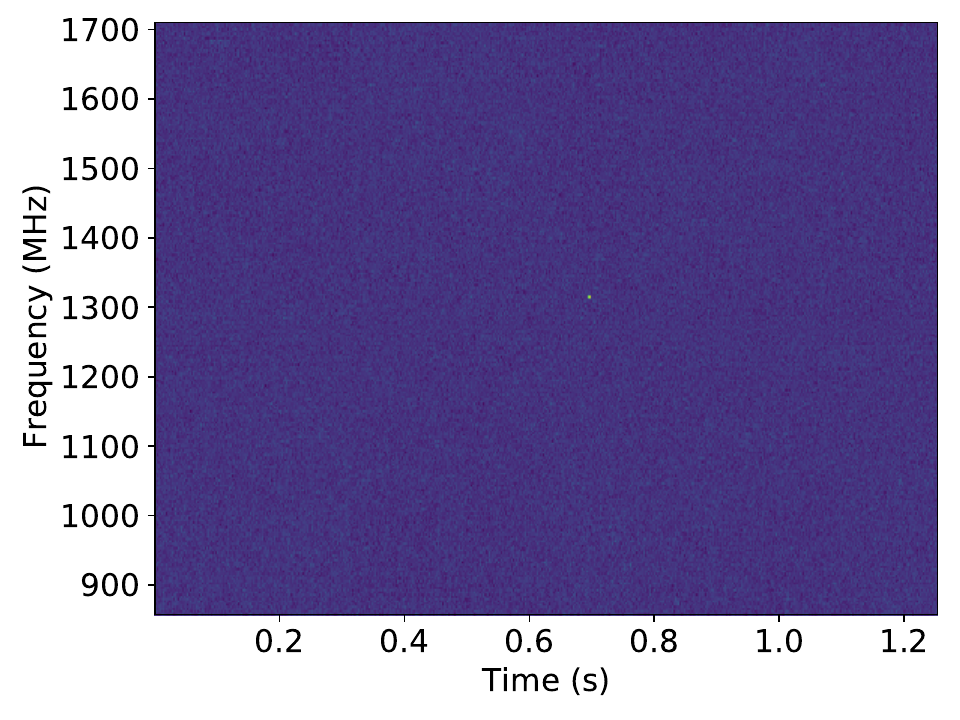}
    \caption{Example of the RFI mitigation with SKF, ZDMF and KF. The test data used is from MMGPS-L and has 2048 frequency channels with a time resolution of 153 us. The SKF, ZDMF, KF are applied successively to the raw data. The top left panel displays the dynamic spectrum of the raw filterbank data. The top right panel shows the dynamic spectrum of the data after SKF, which removes the bad frequency channels. The bottom left panel shows the dynamic spectrum of the data after ZDMF, which eliminates zero-DM RFI signals and the baseline variation. Finally, the bottom right panel shows the dynamic spectrum of the data after KF, which removes the RFI signals with long duration.}
    \label{fig:rfi}
\end{figure*}

RFI signals are a common problem in radio observations. If not mitigated, RFI signals can significantly affect the profiles estimated during the folding process. Some algorithms have been widely used to mitigate the RFI signals \citep[e.g.][]{Offringa2012A&A, Men2019MNRAS, Morello2022MNRAS}. In our program, there are several RFI mitigation algorithms available, including the skewness-kurtosis filter (SKF), the zero-DM matched filter (ZDMF), and the Kadane filter (KF): (1) The SKF algorithm removes outliers based on the skewness and kurtosis of time samples in different frequency channels; (2) The ZDMF algorithm removes the correlated component between frequency channels in each channel, as discussed in \citet{Men2019MNRAS}; (3) The KF algorithm searches for the maximum summation along time samples in each frequency channel based on Kadane's algorithm and replaces them by the mean value or a random value if the S/N is beyond a given threshold. \FIG{rfi} shows an example of the SKF, ZDMF, and KF algorithms tested on observation data of MMGPS-L. Previous research has proposed using kurtosis-based RFI mitigation in baseband data \citep{Nita2007PASP}. In the SKF, we extend this approach to filterbank data. Due to the presence of non-Gaussian noise in filterbank data, we adopt a threshold based on the inter-quartile range (IQR) instead, that is similar to the Inter-Quartile Range Mitigation (IQRM) approach presented in \cite{Morello2022MNRAS}. The skewness and kurtosis definitions of SKF are shown in \APP{skewness_kurtosis}, and the S/N definition of KF is shown in \APP{KF}.

\subsection{Dedispersion}
\label{sec:dedispersion}
High frequency resolution is unnecessary for candidate classification. In addition, dedispersing the data into sub-bands with low frequency resolution can accelerate the folding process by reducing the number of channels to be processed. As discussed in \SEC{Benchmark}, dedispersion becomes the bottleneck of the entire pipeline, and we have adopted the Fast Discrete Dispersion Measure Transform (FDMT) algorithm \citep{Zackay2017ApJ} to speed up this process. However, the FDMT algorithm is designed for equally spaced DM trials, while candidate DMs are typically unevenly spaced. To address this, we have modified the FDMT algorithm  with a pruning strategy to accommodate the large, non-uniform DM trials encountered in dedispersion. The modifications consist of three parts: (1) Reordering the DM trials to enable in-place computation, which improves cache-friendliness. (2) Pruning the unused intermediate dedispersion computations. (3) Applying partial iterations of the FDMT algorithm to obtain the sub-band data. We refer to our modified algorithm as the "pruned Fast Discrete Dispersion Measure Transform" (pFDMT). Since we are working with filterbank power data, we apply incoherent dedispersion to correct for dispersion delays between frequency channels. The delay between frequencies $f_1$ and $f_2$ can be expressed as
\begin{equation}
     \tau_d = \frac{e^2}{2\pi m_e c} \RM{DM} \left(\frac{1}{f_1^2}-\frac{1}{f_2^2}\right)\,,
     \label{eq:dispersion_delay}
 \end{equation}
where $e,m_e,c$ are the elementary charge, electron mass and speed of light in a vacuum, respectively.

The pFDMT algorithm consists of two steps: (1) Generating the dedispersion tree, which utilizes the FDMT algorithm with a cache-friendly design that allows for in-place computation, and (2) Pruning the tree to reduce the processing. To help explain the pFDMT algorithm, we first provide some definitions:
\begin{itemize}
    \item \emph{node}: A time series of one frequency channel with a specific DM.
    \item \emph{attribute}: A parameter pair, consisting of a DM and channel index.
    \item \emph{atomic dedispersion}: The process of dedispersing two adjacent channels with one DM.
    \item \emph{butterfly}: A butterfly is composed of two consecutive atomic dedispersions performed on adjacent channels with two successive DMs.
    \item \emph{stage}: Each stage consists of multiple butterflies.
    \item \emph{root}: A final dedispersed time series with all frequency channels scrunched of a particular DM in the final stage.
\end{itemize}
In this paragraph, we will describe the FDMT algorithm, while the modification will be described in the following paragraphs. The dedispersion tree is composed of multiple stages, where each stage consists of multiple butterflies. In each stage, the butterflies perform dedispersion on adjacent channels using a finer DM grid compared to the previous stage. This transformation results in new nodes with a DM step and a number of channels that are halved compared to the previous stage after each stage. Initially, the data consists of $N_f$ frequency channels, at a single DM trial, e.g. DM=0. In each step, pairs of neighbouring channels are dedispersed at 2 DM trials spaced by half the total DM range, and summed together. After the $n$-th stage, there are therefore $N_f/2^n$ sub-bands, dedispersed at $2^n$ DM values. As a result, the number of stages can be calculated as $\log_2 N_f$. In the first stage, each pre-transformed node's attributes correspond to the channel index and the same DM of the starting DM trial. In the final stage, we obtain the dedispersed time series of the DM trials.

To reduce the algorithm's complexity, unused atomic dedispersions can be pruned using the following steps: (1) Starting from the nodes after the final stage and identifying the unused roots with DMs that are not among the candidate DMs; (2) Pruning the atomic dedispersions that lead to these roots; (3) Moving to the previous stage and identifying nodes that do not have atomic dedispersions to the pre-transformed nodes in the next stage; (4) Pruning these atomic dedispersions; (5) Repeating steps (3) and (4) until the first stage is reached.

To reduce the smearing caused by DM errors, we can adjust the DM step in the final stage of our algorithm. Furthermore, to improve the range of dedispersed DMs, we can apply the pFDMT algorithm to multiple DM trials with different starting DM values. For larger DMs, intra-channel smearing becomes significant, and we can enhance computational efficiency by integrating samples to a coarser time resolution and increasing the DM step.

\begin{figure*}
    \centering
    \includegraphics[width=1.8\columnwidth]{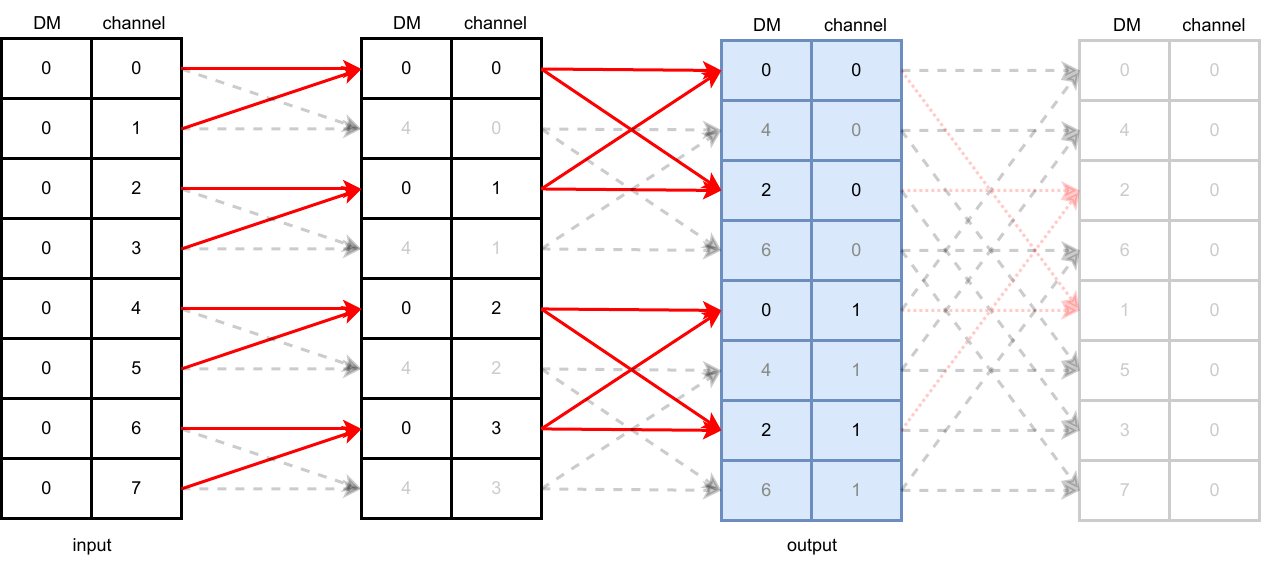}
    \caption{Diagram illustrating an example of the pFDMT algorithm using 8 frequency channels. The dedispersion is only applied to $\RM{DM}=[1,2]$, resulting in a dedispersed sub-band data of 2 frequency channels. The output data is filled with light blue. The start DM is 0 and the DM step is 1, which can both be adjusted. The red solid line indicates that the "atomic dedispersion" is non-pruned, while the black transparent dashed line or red transparent dotted line indicate that it is pruned. The red transparent dotted line represents the pruned "atomic dedispersion" in the final stage which is not used, as only sub-band data is needed.}
    \label{fig:pFDMT}
\end{figure*}

Since we only require the sub-band data and not the full dedispersed time series, we can optimize the algorithm by stopping at the intermediate stage that corresponds to the desired number of frequency channels, denoted as $N_\RM{sub}$. The computational operations of the pFDMT algorithm can be expressed as $\eta N_t N_f \log_2 \frac{N_f}{N_\RM{sub}}$, where $N_t$ and $N_f$ represent the number of time samples and frequency channels, respectively. The filling factor of all atomic dedispersions, denoted by $\eta$, is dependent on the distribution of the DM trials. In contrast, the operations of performing brute-force dedispersion for each candidate are $N_t N_f N_\RM{cand}$, where $N_\RM{cand}$ is the number of candidates. The computational operations are reduced by a factor of $\frac{N_\RM{cand}}{\eta \log_2 (N_f/N_\RM{sub})}$. The pseudo-code of the pFDMT algorithm is presented in \AG{pFDMT}, and \FIG{pFDMT} illustrates an example with eight frequency channels. To execute the pFDMT algorithm in a CPU cache-friendly manner, we implement a depth-first traversal recursive process, and the butterfly can be performed in-place. Hence, the space complexity of the algorithm is $\mathcal{O}(N_t N_f)$.

\begin{algorithm}
	\caption{Sparse Tree Dedispersion Transform. $\pmb{D}$ is a 1-dimensional array with the shape of ($N_f$), and $\pmb{D}_{idm, ichan} = \pmb{D}[ichan * ndm + idm]$, where $ndm$ is the number of DMs at the current stage. $\pmb{X}$ is the spectra before performing the first stage, e.g. a 2-dimensional array with the shape of ($N_f$, $N_t$), which is updated in the following stages. Specifically, $\pmb{X}_{idm, ichan} = \pmb{X}[ichan * ndm + idm, :]$. $depth$ is defined as the difference in stage indices between the root and the current stage of the dedispersion tree.}
	\begin{algorithmic}[1]
	    \For {$isub$ = 1, 2, \ldots, $nsub$}
	        \State  \Call{dedisperse}{$depth_{nsub}$, $isub$}
	    \EndFor
	    
		\Procedure{dedisperse}{$depth$, $ichan$}
		    \If {$depth$ == $\log_2 N_f$}
		        \State $\pmb{D}_{0,ichan}$ = DM start
		        \State \Return
	        \EndIf
	        
	        \State \Call{dedisperse}{$depth$+1, 2*$ichan$}
	        \State \Call{dedisperse}{$depth$+1, 2*$ichan$+1}
	        
	        \For {$idm=1,2,\ldots,ndm$}
	            \State $\pmb{D}_{idm,2*ichan+1}$ += DM step at current depth, i.e. $2^{depth}$ times DM step at the root
	            \State $d_0$ = $\pmb{D}_{idm,2*ichan}$
	            \State $d_1$ = $\pmb{D}_{idm,2*ichan+1}$
	            \State \Call{butterfly}{$\pmb{X}_{idm,2*ichan}$, $\pmb{X}_{idm,2*ichan+1}$, $d_0$, $d_1$}
	        \EndFor
	    \EndProcedure
	    
	    \Procedure{butterfly}{$\pmb{x}$, $\pmb{y}$, $d_0$, $d_1$}
		    \If {path0 is activated}
		        \State perform dedispersion between $\pmb{x}$ and $\pmb{y}$ with DM = $d_0$  
		        \State update $\pmb{x}$ with the dedispersed time series above
		    \EndIf
		    
		    \If {path1 is activated}
		        \State perform dedispersion between $\pmb{x}$ and $\pmb{y}$ with DM = $d_1$  
		        \State update $\pmb{y}$ with the dedispersed time series above
		    \EndIf
	    \EndProcedure
	\end{algorithmic} 
	\label{ag:pFDMT}
\end{algorithm}

\subsection{Folding}
To enhance the visual significance of the candidate's profile and preserve its time variation information, we can perform intensity integration of the sub-banded data based on the spin phase over successive time spans. This process, known as folding, can be viewed as the estimation of the profile $s$ of a periodic signal from the intensity time series $x$. If we disregard the effects of sampling, we can model the intensity $x$ as given by
\begin{equation}
     x(t) = s(\phi(t)) + n(t)\,,
     \label{eq:signal_model}
\end{equation}
where $\phi(t)$ is the phase at time $t$ that can be calculated from \EQ{phase}. $n$ is the noise, which is assumed as Gaussian white noise with a mean of zero and a variance of $\sigma^2$ in our calculation.
The profile $s$ can be approximated by a step function, given by
\begin{align}
     s(\phi) &= \sum_{k=0}^{N-1} c_k f_k(\phi)\,,\label{eq:profile_approx1}\\
     f_k(\phi) &= \left\{
        \begin{array}{ll}
            1 & \quad k<N\phi<k+1\,,\\
            0 & \quad \RM{otherwise}\,,
        \end{array}
    \right.
     \label{eq:profile_approx}
\end{align}
where $k$ represents the $k$th phase bin of the profile, i.e. $f_k(\phi)=1$ when the phase $\phi$ of a sample is located in the phase range of the $k$th phase bin. $N$ represents the number of phase bins. $c_k$ is the coefficient of $f_k(\phi)$, which can be estimated using the least squares method. $\chi^2$ is defined as
\begin{equation}
     \chi^2 = \sum_{i} \frac{\left(x(t_i)-\sum_{k=0}^{N-1} c_k f_k (\phi(t_i))\right)^2}{\sigma^2}\,,
     \label{eq:chisq}
\end{equation}
where $i$ represents the $i$th sample. It can be minimized to obtain the coefficients
\begin{equation}
    c_k = \frac{1}{C_k} \sum_{i,k<N\phi(t_i)<k+1} x(t_i) \,,
    \label{eq:c_k}
\end{equation}
where $C_k$ is the number of samples that locate in the $k$th phase bin. This is the algorithm used in the software package {\sc{DSPSR}} \citep{Straten2011PASA}.

However, the time resolution of the profile estimated from this traditional folding algorithm is limited by the time resolution of the integrated samples in the raw data, as the integration effect introduced by sampling is not considered in \EQ{signal_model}. To address this issue, we propose a novel folding algorithm based on the Tikhonov-regularised least squares method (TRLSM), also known as ridge regression \citep{Tikhonov1943OnTS, Arthur1970Tech}. We modify the signal model \EQ{signal_model}, given by
\begin{equation}
     x(t) = \int_{\phi(t-\Delta t/2)}^{\phi(t+\Delta t /2)} s(\phi(t)) d \phi(t) + n(t)\,,
     \label{eq:signal_model_mod}
\end{equation}
where $\Delta t$ is the time resolution of one sample. Combined with \EQ{profile_approx1} and \EQ{profile_approx}, we have
\begin{equation}
     x(t) = \sum_{k=0}^{N-1} w_k(t) c_k + n(t)\,,
     \label{eq:signal_model_der}
\end{equation}
where $w_k(t)$ is the fraction of the $k$th phase bin swept by the sample at time $t$. Solving $c_k$ using the normal least squares method in \EQ{chisq} is an inverse problem that has a stability problem, which can be handled using the Tikhonov regularised $\chi'^2$, i.e.
\begin{equation}
     \chi'^2 = \sum_{i} \frac{\left(x(t_i)-\sum_{k=0}^{N-1} w_k(t_i) c_k\right)^2}{\sigma^2} + \lambda \sum_{k=0}^{N-1}c_k^2\,,
     \label{eq:tk_chisq}
\end{equation}
where $\lambda$ is the ridge parameter (see later \SEC{ridge_parameter} for a discussion of that). \EQ{tk_chisq} can be represented in the matrix form, i.e.
\begin{equation}
     \chi'^2 = \frac{(\pmb{x}-\pmb{W} \pmb{c})^T (\pmb{x}-\pmb{W} \pmb{c})}{\sigma^2} + \lambda \pmb{c}^T \pmb{c}\,,
     \label{eq:tk_chisq_mx}
\end{equation}
where $T$ represents the transposition. Minimising $\chi'^2$ gives
\begin{equation}
    \pmb{c} = (\pmb{W}^T \pmb{W}/\sigma^2 + \lambda \pmb{I})^{-1} \pmb{W}^T \pmb{x} / \sigma^2 \,,
    \label{eq:coeff}
\end{equation}
where $\pmb{I}$ is the identity matrix.

\begin{figure}
    \centering
    \includegraphics[width=\columnwidth]{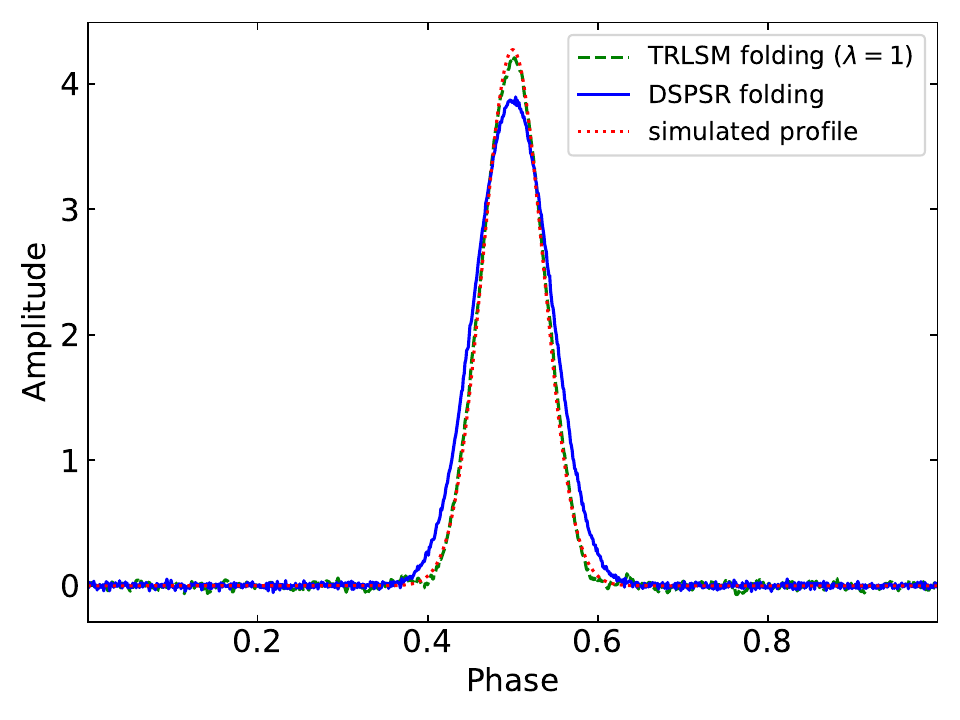}
    \caption{Comparison between the TRLSM and DSPSR folding algorithms, using a folding example. The simulated profile is represented by a red dotted line, while the profile estimated from the DSPSR folding algorithm is shown in blue solid line. The TRLSM folding algorithm is represented by a green dashed line, with a ridge parameter of $\lambda=1$.}
    \label{fig:trlsm}
\end{figure}

To investigate the performance of the TRLSM folding algorithm, we conducted a folding test on simulation data that includes a periodic signal with Gaussian white noise. The signal has a Gaussian profile with a frequency of 650 Hz, and the data has a time resolution of 130 $\mu$s. \FIG{trlsm} illustrates the improvement of profile resolution achieved with the TRLSM folding algorithm compared to the DSPSR folding algorithm. Further investigation of this folding algorithm will be presented in future work (Men et al., in prep.).

In most real data folding cases, the time resolution is usually smaller than or comparable to the phase resolution of the profile. Therefore, the weight matrix $W$ is a sparse matrix, which significantly reduces the computing complexity of \EQ{coeff}, i.e. $\mathcal{O}(N_\RM{sub} N_t+N_b^3)$, where $N_n$ is the number of phase bins of the profile.

\subsection{DM, ${\nu}$, ${\dot{\nu}}$ optimization}
To find the optimal DM, $\nu$, and $\dot{\nu}$ parameters for each candidate, an additional optimisation step is required as the coarse grid of the parameter space in acceleration search may not be optimal. We propose a novel iterative algorithm in our pipeline consisting of several steps in each iteration: (1) Calculate the integrated time-phase spectrum by scrunching the archive along all frequency channels; (2) Optimize $\nu$ and $\dot{\nu}$ by maximizing the $\chi_s^2$ of the integrated profile, which is calculated by scrunching the time-phase spectrum along time. Here, $\chi_s^2$ is defined as
\begin{equation}
    \chi_s^2 = \sum_{k=0}^{N-1} \frac{(s_i - \bar{s})^2}{\sigma_s^2} \,,
\end{equation}
where $\bar{s}$ and $\sigma_s^2$ are the mean and variance of the noise inferred from the archive; (3) Correct the phase shifts of the profiles in the original archive with the updated $\nu$ and $\dot{\nu}$; (4) Calculate the integrated frequency-phase spectrum by scrunching the updated archive along time; (5) Optimize DM by maximizing $\chi_s^2$ of the integrated profile, which is calculated by scrunching the frequency-phase spectrum along frequency; (6) Correct the phase shifts of the profiles in the updated archive with the updated DM. This iterative procedure terminates when the change of the parameters is less than a predefined precision. We use $\chi_s^2$ instead of S/N as the criterion for the significance of the profile because it can be efficiently computed. The complexity of this iterative optimization algorithm is significantly lower than the brute-force algorithm that searches for the best DM, $\nu$ and $\dot{\nu}$ in a three-dimensional grid.

\section{Benchmark}
\label{sec:Benchmark}

\begin{figure*}
    \centering
    \includegraphics[width=\columnwidth]{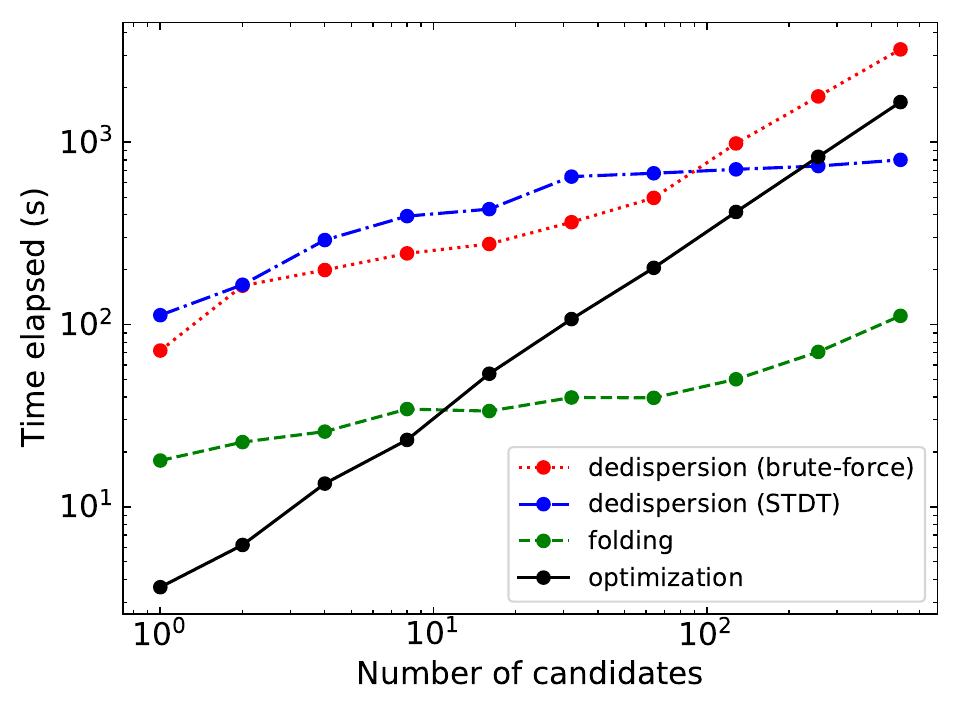}
    \includegraphics[width=\columnwidth]{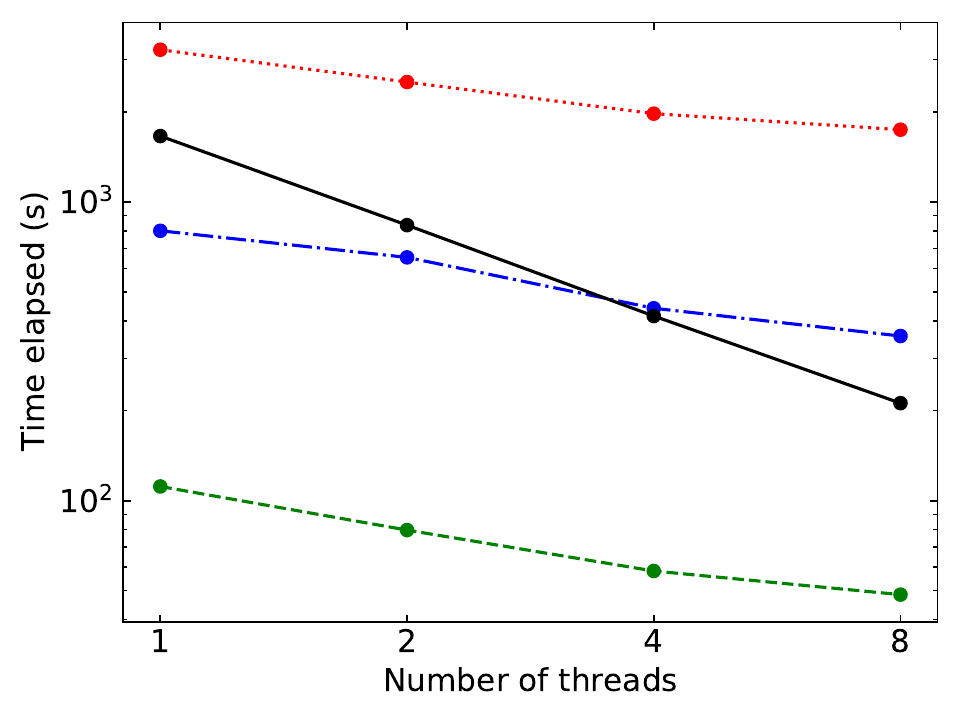}
    \caption{Benchmark results of the folding pipeline, which includes the dedispersion, folding and optimization process. The left panel illustrates the relation between time consumption and the number of candidates running on a single thread. The right panel shows the relation between time consumption and the number of threads with 512 candidates. The red dotted line represents the brute-force dedispersion, while the blue dash-dot line represents the pFDMT algorithm. The folding process is represented by the green dashed line, and the optimization process is represented by the black solid line.}
    \label{fig:benchmark}
\end{figure*}

To evaluate the performance of our folding pipeline, we generated a simulated data set with a similar format as the data in MMGPS-L, featuring a time resolution of 153 $\mu$s, 2048 frequency channels, and 10 minutes observation time. Rather than performing a full acceleration search, we instead simulated the input parameters of candidates for the folding pipeline. We used a logarithmic uniform distribution for the spin frequency ranging from 0.1 to 1000 Hz and a fixed zero value for the spin frequency derivative, which did not affect the folding and optimization performance. We also used an equally spaced DM grid from 0 to 3000 \cmpc{}, which is the worst case for the pFDMT algorithm.

In \textsc{PulsarX}, there are two implementations of the folding pipelines, i.e. \texttt{psrfold\_fil} and \texttt{psrfold\_fil2}. The only difference between them is the dedispersion algorithm they employ. The former utilizes brute-force dedispersion, while the latter utilizes the pFDMT algorithm. We benchmarked the folding pipelines on a {\sc{Intel(R) Xeon(R) Silver 4116 CPU @ 2.10GHz}}, which is used in MMGPS-L. The results are presented in \FIG{benchmark}. From the left panel, we can see that the pFDMT algorithm becomes more efficient than the brute-force dedispersion algorithm as the number of candidates increases, since the pFDMT algorithm has smaller memory I/O and computing complexity. The folding process consumes only about 10\% of the time, which increases slowly in the regime of fewer candidates and becomes close to linearly scaled with more candidates. This is due to the fact that the memory I/O is reduced when there are more candidates since the folding process will reuse the sub-band dedispersed data. This can also explain that the time consumption is not linearly scaled with the number of threads, as shown in the right panel of \FIG{benchmark}. The time consumption on the optimization process is linearly scaled with the number of candidates as expected.  It is expected that the optimization process will dominate the time consumption when the number of candidates becomes larger. The right panel of \FIG{benchmark} shows the scale relation of the time consumption with the number of threads, which is not linearly scaled except for the optimization process, because the performance of the other processes are limited by memory I/O rather than computing power. In conclusion, \texttt{psrfold\_fil2} can operate almost in real-time with 8 CPU threads for about 500 candidates, which is highly efficient.

\section{Discussions}
\label{sec:Discussions}
\subsection{The ridge parameter $\lambda$}
\label{sec:ridge_parameter}
In the TRLSM folding algorithm, the ridge parameter $\lambda$ is adjustable. Choosing a small value for $\lambda$ can result in a noisy profile, while selecting a large value can lead to broadening of the profile and reduced resolution. Finding the optimal ridge parameter for different regimes has been extensively studied \citep{Ayinde2016HJMS}. In our case, we can optimize $\lambda$ to enhance the S/N of the profile. Here, we present a brief principle solution, and more details will be provided in future work (Men et al., in prep.). Firstly, we can demonstrate the algorithm within the Bayesian framework, where we interpret $\lambda$ as the prior precision (i.e., the inverse variance) of the pulse profile amplitude for a single phase bin. In this framework, the likelihood and prior can be given as
\begin{align}
    P(\pmb{x}|\pmb{c}) &= \frac{1}{(2 \pi \sigma^2)^{N/2}} \exp\left(-\frac{1}{2 \sigma^2} (\pmb{x}-\pmb{W} \pmb{c})^T (\pmb{x}-\pmb{W} \pmb{c})\right) \,, \\
    P(\pmb{c}) &= \frac{1}{(2 \pi/\lambda)^N/2} \exp\left(-\frac{\lambda}{2} \pmb{c}^T \pmb{c} \right) \,,
\end{align}
respectively. We can then define the Bayes factor as
\begin{equation}
    K = \frac{\int P(\pmb{x}|\pmb{c}) P(\pmb{c}) d \pmb{c}}{P(\pmb{x}|\pmb{c=0})}\,.
\end{equation}
By maximising $K$, which maximises the chi-square of the pulse profile, but with an "Occam's razor" penalty factor that prevents over-fitting (noisy pulse profiles) caused by a very small $\lambda$, we can obtain the optimal ridge parameter $\lambda$.

\begin{figure}
    \centering
    \includegraphics[width=\columnwidth]{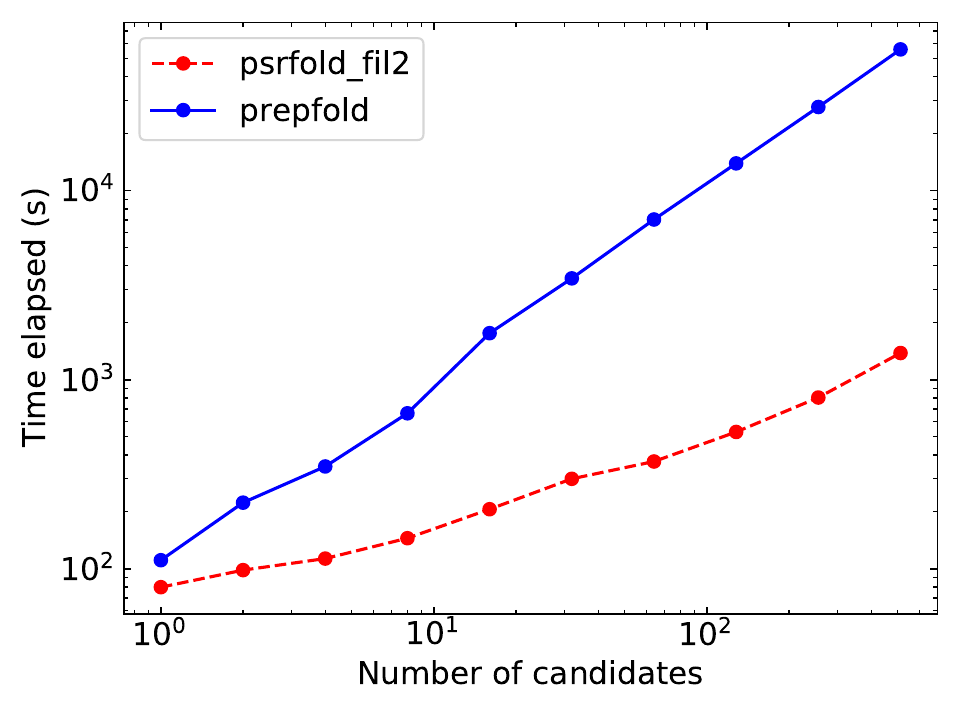}
    \caption{Benchmark results obtained from a 10-minute simulation dataset with 2048 frequency channels and a time resolution of 153 $\mu$s, comparing the folding program \texttt{psrfold\_fil2} in {\sc{PulsarX}} with \texttt{prepfold} in {\sc{PRESTO}}. The benchmark was conducted using an Intel(R) Core(TM) i7-10750H CPU clocked at 2.60GHz, along with an NVMe SSD boasting a high bandwidth of approximately 2 GB/s.}
    \label{fig:presto_pulsarx}
\end{figure}

\subsection{Comparison with \sc{PRESTO}}
{\sc{PRESTO}\footnote{https://github.com/scottransom/presto.git}} is a widely used pulsar search package that offers various tools for different tasks, including \texttt{rfifind} for RFI mitigation, \texttt{prepsubband} for dedispersion, \texttt{prepfold} for candidate folding, and so on. In this context, we will provide a brief comparison between \texttt{prepfold} and \texttt{psrfold\_fil2}, which both perform similar tasks, including RFI mitigation, dedispersion, folding, and optimization, but with different algorithms. \texttt{prepfold} performs these tasks for each candidate in a single run and employs a brute-force dedispersion approach for each candidate. In the folding process, \texttt{prepfold} estimates the profile using $\pmb{c} = \pmb{W}^T \pmb{x}$, i.e. $\lambda = \infty$ in \EQ{coeff} \citep{Bachetti2021ApJ}. Additionally, \texttt{prepfold} applies brute-force optimization in the three-dimensional grid of DM, $\nu$, and $\dot{\nu}$. Furthermore, \texttt{prepfold} has the capability to fold intermediate products of dedispersion in the searching stage, such as the dedispersed time series or sub-band spectrum. It can perform the conversion of time from topocentric to barycentric reference frame, along with folding the binary candidate using the orbital parameters. These features are not currently supported in \texttt{psrfold\_fil2}. However, efficient raw data folding can save the storage and disk I/O by eliminating the need for intermediate products of dedispersion in the searching stage. To improve the efficiency of the folding pipeline, \texttt{psrfold\_fil2} applied several different algorithms, including (1) folding multiple candidates simultaneously to reduce the disk I/O bandwidth; (2) using the pFDMT algorithm to speed up dedispersion before folding when there are many candidates; (3) using a novel iterative optimization algorithm to speed up the DM,$\nu$,$\dot{\nu}$ optimization significantly compared to the brute-force optimization in a three-dimensional grid. We conducted a real benchmark comparison between the folding program in {\sc{PulsarX}} and {\sc{PRESTO}}, as depicted in \FIG{presto_pulsarx}. The results reveal that \texttt{psrfold\_fil2} outperforms \texttt{prepfold} by approximately 50 times when dealing with a large number of candidates. This benchmark employed an NVMe SSD with an I/O bandwidth of around 2 GB/s. However, it's worth noting that \texttt{prepfold} demands significantly greater disk I/O bandwidth, which implies that its performance will be notably slower when executed on a Hard Disk Drive (HDD). These features make \texttt{psrfold\_fil2} particularly suitable for handling the folding processing of pulsar surveys with a large data rate, e.g. MMGPS and TRAPUM.

\begin{figure}
    \centering
    \includegraphics[width=\columnwidth]{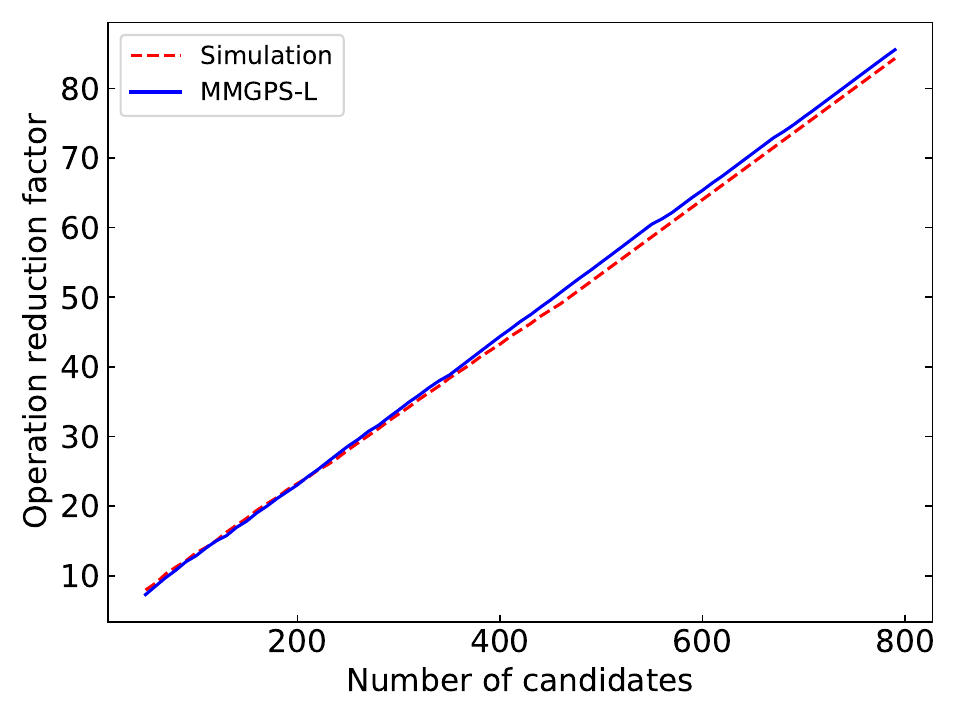}
    \caption{Operation reduced factor achieved by the pFDMT algorithm compared to the brute-force algorithm. The blue solid line represents the test data obtained from MMGPS-L, where the number of candidates is determined based on the S/N cutoff. On the other hand, the red dashed line represents the simulated data with an equally spaced distribution of DM.}
    \label{fig:pFDMT_imfactor}
\end{figure}

\begin{figure*}
    \centering
    \includegraphics[width=1.6\columnwidth]{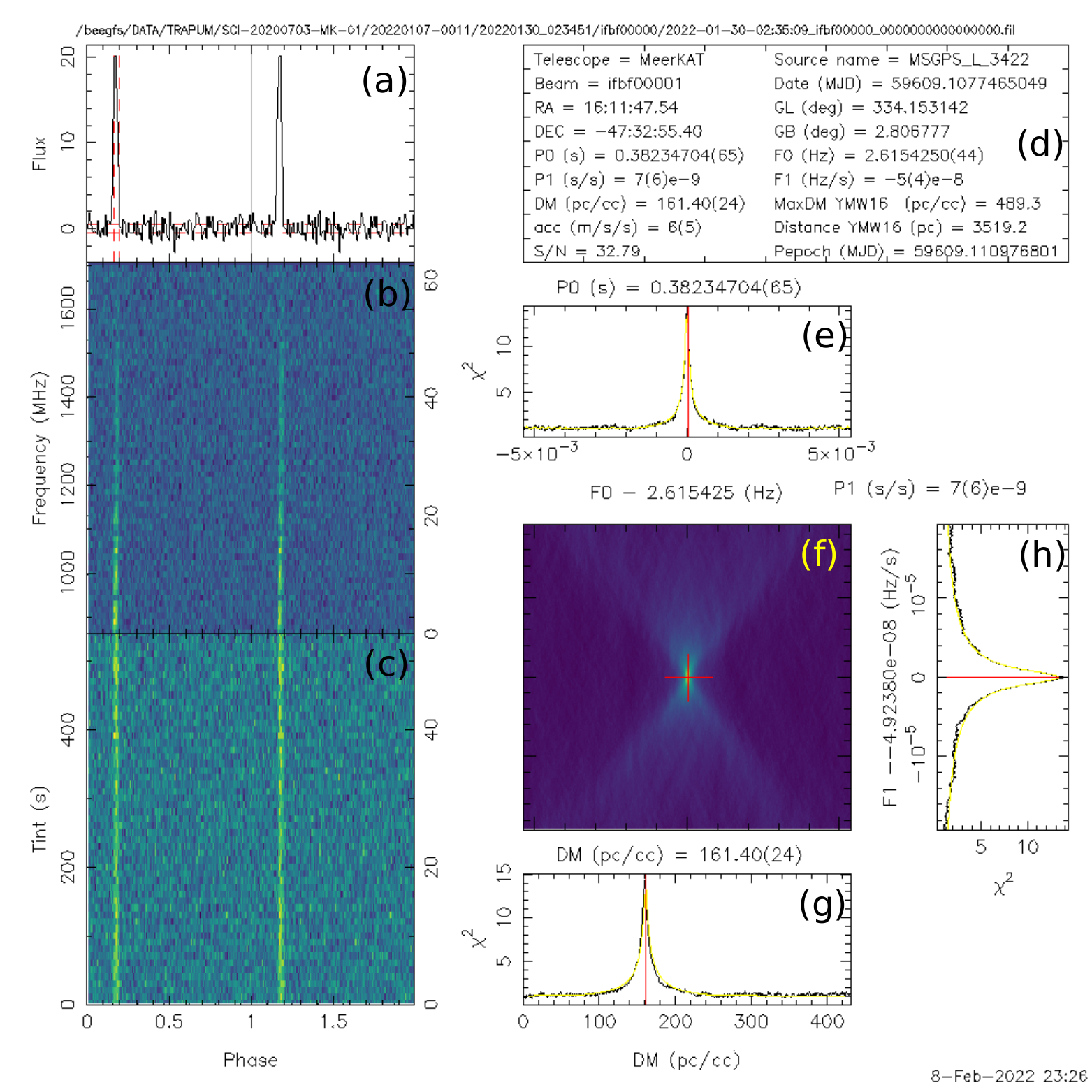}
    \caption{Candidate plot generated by the folding pipeline of MMGPS-L, showing a known pulsar, PSR B1609-47. The plot is divided into 8 panels: Panel (a) shows the folded profile of a candidate, with a grey vertical span indicating intra-channel dispersion smearing and red lines indicating pulse and noise amplitudes; Panel (b) shows the frequency-phase spectrum with time spans integrated; Panel (c) shows the time-phase spectrum with frequency channels integrated; Panel (d) shows the meta information of the observation and candidate; Panel (e) shows the $\chi^2$-$\nu$ relation; Panel (f) shows the $\chi^2$ spectrum of the $\nu$-$\dot{\nu}$ plane, with a red cross representing the pre-optimized $\nu$ and $\dot{\nu}$; Panel (g) shows the $\chi^2$-DM relation; Panel (h) shows the $\chi^2$-$\dot{\nu}$ relation. In panels (e), (g) and (h), the red line represents the pre-optimized value, and the yellow line shows the fitting curve of the relation under the assumption that the candidate's profile is Gaussian with a pulse width estimated from the folded profile.}
    \label{fig:mmgps}
\end{figure*}

\subsection{Application on the MMGPS-L}
{\sc{PulsarX}} has been employed within the MMGPS-L and TRAPUM projects. The TRAPUM initiative involves multiple sub-projects, including the Globular Cluster Survey, Fermi Source Survey, Nearby Galaxy Survey, and more, each characterized by distinct configurations. In this context, we will focus on elucidating the specific application of {\sc{PulsarX}} within the MMGPS-L project. The MMGPS-L is a wide field-of-view pulsar survey conducted with the MeerKAT telescope. Detailed information about the MMGPS-L configuration can be found in \cite{Padmanabh2023arXiv}. This survey generates approximately 21 TB of data per hour, making it crucial to develop a high-performance processing pipeline capable of handling such a high data rate while working within the limitations of available disk space. For acceleration search, we utilize the GPU-based acceleration search package, {\sc{PEASOUP}}, while candidate folding is performed using {\sc{PulsarX}}. As {\sc{PulsarX}} is CPU-based, it can run simultaneously with {\sc{PEASOUP}} without resource competition. The time required for candidate folding in the MMGPS-L processing can be accommodated within the duration of the acceleration search. \FIG{pFDMT_imfactor} illustrates an example of the operation reduction factor achieved by the pFDMT algorithm using real DM trials of the candidates from a MMGPS-L observation. Notably, it's close to the equal-spaced DM trials. From the information presented in Figure \FIG{pFDMT_imfactor}, it can be observed that the pFDMT algorithm achieves a significant operation reduction factor of approximately 50 when processing around 500 candidates. \FIG{mmgps} shows an example plot generated by the MMGPS-L processing pipeline.

\section{Conclusions}
With the era of the Square Kilometer Array (SKA) approaching, more powerful computing hardware will be required to handle the massive amounts of data generated. Nonetheless, it is still worth exploring the more efficient algorithms that can significantly enhance the performance of the software we use. Our work introduced a new high performance folding program that includes an efficient dedispersion algorithm and parameter optimization, speeding up the pulsar searching pipeline significantly. Additionally, we developed a skewness and kurtosis-based RFI mitigation algorithm to remove the frequency channels contaminated by RFI signals. A novel folding algorithm is proposed that can improve the resolution of the profile estimated from the folding process. We also demonstrate the application of the program on MMGPS, showcasing its efficiency in handling the high data rate of this wide field-of-view interferometer-based pulsar survey. This provides inspiration for improving the performance of the pulsar searching pipeline in the future radio telescopes, e.g. MeerKAT+ and SKA.
\label{sec:Conclusions}

\begin{acknowledgements}
The authors would like to thank Scott Ransom for his helpful discussion regarding PRESTO.

The MeerKAT telescope is operated by the South African Radio Astronomy Observatory, which is a facility of the National Research Foundation, an agency of the Department of Science and Innovation. SARAO acknowledges the ongoing advice and calibration of GPS systems by the National Metrology Institute of South Africa (NMISA) and the time space reference systems department of the Paris Observatory.

TRAPUM observations used the FBFUSE and APSUSE computing clusters for data acquisition, storage and analysis. These clusters were funded and installed by the Max-Planck-Institut für Radioastronomie and the Max-PlanckGesellschaft.

EC acknowledges funding from the United Kingdom's Research and Innovation Science and Technology Facilities Council (STFC) Doctoral Training Partnership, project reference 2487536. 

YPM, EB and GD acknowledge continuing support from the Max Planck society. 
\end{acknowledgements}

\bibliography{ms}

\begin{appendix}

\section{skewness and kurtosis}
\label{sec:skewness_kurtosis}

The skewness $\gamma$ and kurtosis $\kappa$ of a time series are defined as
\begin{align}
    \gamma &= \frac{\frac{1}{N} \sum_{i=0}^{N-1} (x_i-\mu)^3}{(\frac{1}{N} \sum_{i=0}^{N-1} (x_i-\mu)^2)^{3/2}} \,, \\
    \kappa &= \frac{\frac{1}{N} \sum_{i=0}^{N-1} (x_i-\mu)^4}{(\frac{1}{N} \sum_{i=0}^{N-1} (x_i-\mu)^2)^2} \,, \\
    \mu &= \frac{1}{N} \sum_{i=0}^{N-1} x_i\,,
\end{align}

respectively, where $i$ represents the $i$th sample of the time series, and $N$ is the number of total samples.

\section{S/N definition}
\label{sec:KF}

After the KF has found the maximum summation in a time series $\pmb{x}$, the detection statistic S/N is defined as
\begin{equation}
    \RM{S/N} = \frac{\sum_{i=a}^{b} x_i}{\sqrt{b-a+1} \sigma} \,,
\end{equation}
where $a$ and $b$ are the count of the start sample and end sample of the duration with maximum summation, i.e. a contiguous sub-array in $\pmb{x}$ with the largest sum. $\sigma$ is the standard deviation of the time series.

\end{appendix}

\end{document}